\documentclass[useAMS,usenatbib]{mn2e}
\usepackage{supertabular,wrapfig,placeins}
\usepackage{graphicx,amsmath,amssymb,subfigure,multirow}
\usepackage{marvosym, times}
\usepackage{graphics}
\usepackage{natbib}
\usepackage{amssymb}
\usepackage{amsmath}
\usepackage{euscript}
\usepackage{setspace}
\usepackage{fancyhdr}
\usepackage{afterpage,rotating,url}

\def\simgt{\mathrel{\lower0.6ex\hbox{$\buildrel {\textstyle >}
 \over {\scriptstyle \sim}$}}}
\def\simlt{\mathrel{\lower0.6ex\hbox{$\buildrel {\textstyle <}
 \over {\scriptstyle \sim}$}}}

\newcommand{\gtsim}{\mbox{{\raisebox{-0.4ex}{$\stackrel{>}{{\scriptstyle\sim}}$}}}}
\newcommand{\ltsim}{\mbox{{\raisebox{-0.4ex}{$\stackrel{<}{{\scriptstyle\sim}}$}}}}

\newcommand{\mnras}{MNRAS}
\newcommand{\apjs}{ApJS}
\newcommand{\apj}{ApJ}
\newcommand{\aap}{A\&A}
\newcommand{\aaps}{A\&ASS}
\newcommand{\apjl}{ApJL}
\newcommand{\araa}{ARA\&A}
\newcommand{\aj}{AJ}
\newcommand{\nat}{Nature}
\newcommand{\jcap}{JCAP}
\newcommand{\pasp}{PASP}
\newcommand{\procspie}{SPIE}

\begin{document}

\title[The VIDEO Survey]{The VISTA Deep Extragalactic Observations (VIDEO) Survey\thanks{Based on data products from observations made with ESO Telescopes at the La Silla Paranal Observatory under ESO programme ID 179.A-2006 and on data products produced by the Cambridge Astronomy Survey Unit on behalf of the VIDEO consortium.}}

\author[Jarvis et al.]
{Matt J.~Jarvis$^{1,2,3}$\thanks{Email: matt.jarvis@astro.ox.ac.uk}, D.G.\  Bonfield$^{2}$, V.A.\ Bruce$^{4}$, J.E.\ Geach$^{5}$, K.\ McAlpine$^{6}$, \and R.J.\ McLure$^{4}$, 
E.~Gonz\'alez-Solares$^{7}$, M.\ Irwin$^{7}$, J.\ Lewis$^{7}$, 
A. Kupcu Yoldas$^{7}$, \and S.~Andreon$^{8}$,  N.J.G.\ Cross$^{4}$, J.P.\ Emerson$^{9}$, G.~Dalton$^{1,10}$, J.S.~Dunlop$^{4}$,  S.T.\ Hodgkin$^{7}$, \and O.~Le F{\`e}vre$^{11}$,  M.~Karouzos$^{12}$,  K.\ Meisenheimer$^{13}$, S.\ Oliver$^{14}$, S\, Rawlings$^{1}$, \and C.\ Simpson$^{15}$,
I.\ Smail$^{16}$,  D.J.B.~Smith$^{2}$, M.\ Sullivan$^{1}$,  W.\ Sutherland$^{9}$, S.V.\ White$^{1}$, \and  J.T.L.\ Zwart$^{3}$\\
\footnotesize
$^{1}$Astrophysics, Department of Physics, Keble Road, Oxford, OX1 3RH \\ 
$^{2}$Centre for Astrophysics, Science \& Technology Research Institute, University of Hertfordshire, Hatfield, Herts, AL10 9AB\\
$^{3}$Physics Department, University of the Western Cape, Private Bag X17, Bellville 7535, South Africa\\
$^{4}$SUPA\thanks{Scottish Universities Physics Alliance} Institute for Astronomy, University of Edinburgh, Royal Observatory, Edinburgh EH9 3HJ\\
$^{5}$Department of Physics, McGill University, Ernest Rutherford Building, 3600 rue University, Montr\'eal, Qu\'ebec H3A 2T8, Canada\\
$^{6}$Department of Physics and Electronics, Rhodes University, Grahamstown, 6139, South Africa\\
$^{7}$Institute of Astronomy, Madingley Road, Cambridge CB3 0HA\\
$^{8}$INAF - Osservatorio Astronomico di Brera, via Brera 28, 20121 Milan, Italy\\
$^{9}$Astronomy Unit, School of Physics and Astronomy, Queen Mary University of London, Mile End Road, London, E1 4NS\\
$^{10}$RALSpace, Rutherford Appleton Laboratory, Harwell Oxford, OX11 0QX\\
$^{11}$Laboratoire d’Astrophysique de Marseille, Aix-Marseille Universit\'e, Marseille, France \\
$^{12}$CEOU-Seoul National University, Republic of Korea\\
$^{13}$Max-Planck-Institut f\"ur Astronomie, K\"onigstuhl 17, D-69117 Heidelberg, Germany\\
$^{14}$Astronomy Centre, Department of Physics and Astronomy, University of Sussex, Brighton BN1 9QH\\
$^{15}$Astrophysics Research Institute, Liverpool John Moores University, Twelve Quays House, Egerton Wharf, Birkenhead, CH41 1LD\\ 
$^{16}$Institute for Computational Cosmology, Durham University, South Road, Durham, DH1 3LE, UK\\
\\
\\
\\
\\
\\
\\
\\
\\
\\
\\
\\
\\
\\
\\
\\
\\
\\
\\
\\
\\
\\
\\
\\
\\
\\
\\
\\
\\
\\
\\
\\
\\
\\
\\
\\
\\
}

\maketitle

\begin{abstract}
In this paper we describe the first data release of the 
the Visible and Infrared Survey Telescope for Astronomy (VISTA) Deep Extragalactic Observations (VIDEO) survey. VIDEO is
a $\sim~12$~degree$^{2}$ survey in the near-infrared $Z$,$Y$,$J$,$H$ and $K_{\rm s}$ bands, specifically designed
to enable the evolution of galaxies and large structures to be traced as a
function of both epoch and environment from the present day out to
z=4, and active galactic nuclei (AGN) and the most massive galaxies up to and into the epoch of reionization. With its depth and area, VIDEO will be able to fully explore
the period in the Universe where AGN and starburst
activity were at their peak and the first galaxy
clusters were beginning to virialize. VIDEO therefore offers a unique
data set with which to investigate the interplay between AGN,
starbursts and environment, and the role of feedback at a time when it was potentially most crucial.  

We provide data over the VIDEO-XMM3 tile, which also covers the Canada-France-Hawaii-Telescope Legacy Survey Deep-1 field (CFHTLS-D1). The released VIDEO data reach a $5\sigma$ AB-magnitude depth of $Z=25.7$, $Y=24.5$, $J=24.4$, $H=24.1$ and $K_{\rm s}=23.8$ in 2~arcsec diameter apertures (the full depth of $Y=24.6$ will be reached within the full integration time in future releases). The data are compared to previous surveys over this field and we find good astrometric agreement with the Two-Micron All Sky Survey, and source counts in agreement with the recently released UltraVISTA survey data.
The addition of the VIDEO data to the  CFHTLS-D1 optical data increases the accuracy of photometric redshifts and significantly reduces the fraction of catastrophic outliers over the redshift range $0<z<1$ from 5.8 to 3.1 per cent in the absence of an $i-$band luminosity prior. However, we expect the main improvement in photometric redshifts will come in the redshift range $1<z<4$ due to the sensitivity to the Balmer and 4000\AA\, breaks provided by the near-infrared VISTA filters. All images and catalogues presented in this paper are publicly available through ESO's “phase 3” archive and the VISTA Science Archive.

\end{abstract}


\section{Introduction}\label{sec:intro}

We are already at a point where we have excellent constraints on
the spatial distribution and properties of galaxies in the local
Universe from the largest surveys ever undertaken, namely the
2-degree field Galaxy Redshift Survey \citep[2dFGRS; ][]{Colless2001}, and the Sloan Digital Sky Survey
\citep[SDSS; ][]{Adelman-McCarthy2008} in the optical, the Two-Micron All Sky Survey \citep[2MASS; ][]{2mass} in the near-infrared, and the IRAS redshift survey \citep{Saunders2000} selected in the
far-infrared.  Over the past few years these surveys to study the formation and evolution of galaxies in the Universe, and how they trace the large-scale structure of the Universe, have been complemented by deeper and narrower surveys at all wavelengths, carefully balancing depth and area to ensure that representative volumes of the Universe are observed.
Most notably the Galaxy and Mass Assembly \citep[GAMA; ][]{Driver2011} Survey covering $\sim 100$ square degrees is building on the legacy of the SDSS and 2dfGRS spectroscopic surveys to obtain accurate redshift information down to an $r-$band limit of $r \sim 19.8$ (corresponding to $z$~$\ltsim\,0.5$), with the UKIRT Infrared Deep Sky Survey Large-Area Survey \citep[UKIDSS-LAS; ][]{Lawrence2007} providing important near-infrared data. At the other extreme, the deepest ground-based surveys such as COSMOS \citep{Scoville2007} and the UKIDSS Ultra Deep Survey coupled with the Subaru-XMM Deep Survey \citep[e.g.][]{Foucaud2007, Furusawa2008}, cover $\sim 1$~degree$^{2}$ fields, relying on photometric redshifts to trace the evolutionary history of galaxies \citep[e.g.][]{Ilbert2009}, coupled with extremely time intensive spectroscopic follow-up\citep[e.g.][]{Lilly2007, lefevre2005}.  

The Visible and Infrared Survey Telescope for Astronomy \citep[VISTA;][]{Emerson2004} is now taking the surveys carried out within UKIDSS one step further, with its larger field of view and improved sensitivity at the bluer ($Z, Y, J$) wavelengths of the near-infrared window. VISTA is undertaking several near-infrared public surveys, one of which is the VISTA Deep Extragalactic Observations (VIDEO) Survey.
The aim of the VIDEO survey is to gain a 
representative view of the Universe at $0< z < 4$, allowing galaxy evolution to be traced over
the majority of the history of the Universe, from the richest clusters to the
field.

The VIDEO survey fields are chosen to incorporate current and future multi-wavelength
data sets to facilitate the broadest exploitation of the new near-infrared survey
data both within ESO and across the globe. The main scientific aims of VIDEO are described below. Throughout this paper we use the AB-magnitude system \citep{OkeGunn1983} and adopt the concordance cosmology $\Omega_{M} = 0.3$, $\Omega_{\Lambda} = 0.7$ and $H_{\circ} = 70$\,km\,s$^{-1}$\,Mpc$^{-1}$.

\section{Key Science Aims}


How and when were massive galaxies formed? When did they assemble the
bulk of their stellar mass and how? Where does this mass assembly
occur?  These are crucial questions which VIDEO is designed to answer.

\subsection{Galaxy formation over the epoch of activity}

The advent of deep multi-wavelength surveys has led to huge
progress in this field. 
There are two ways to measure the build-up of stellar mass: by directly
observing star formation, or by measuring the total stellar mass already
formed by a given epoch. The first technique was revolutionised by the
Lyman break technique \citep[e.g.][]{Guhathakurta1990,Steidel1996} which found large numbers
of star-forming galaxies at $z \sim 3$ and GALEX has stretched this to lower redshifts
\cite[e.g.][]{Martin2005}. However, these techniques are only
sensitive to relatively unobscured systems, and are known to miss much
of the luminosity density arising from galaxies at $z\gtsim\,1$\citep[e.g.][]{Schim2005,Magnelli2009}. The {\em Herschel Space Observatory} is now opening up our view of the Universe with both wide- \cite[e.g.][]{Eales2010} and deep-field \cite[e.g.][]{Hermes} surveys, which together probe sufficient volume over a large fraction of cosmic time to determine the full spectral energy distribution of galaxies \cite[e.g.][]{Smith2012} and also the evolution of dust and obscured star formation in the Universe \cite[e.g.][]{Dunne2011,Wuyts2011,Lapi2011}.

On the other hand, mass estimates of distant galaxies require
rest-frame near-infrared wavelength observations to study the parts of galaxy spectral energy distributions that are
not dominated by young stars with low mass-to-light ratios. Surveys
with ground-based near-infrared observations have allowed substantial recent progress in
this area \citep[e.g.][]{Cirasuolo2010, Ono2010,Caputi2011}, particularly at the bright end of the luminosity function at $z\sim2$ where both area and moderately deep data are required. Surveys such as the Great Observatories Origins Deep Survey \citep[GOODS; ][]{GOODS} and the Cosmic Assembly Near-infrared Deep Extragalactic Legacy Survey \citep[CANDELS; ][]{Grogin2011,Koekemoer2011} probe
to much greater depths but only cover very small sky areas, resulting in studies concentrating on the faint galaxy populations \citep[e.g.][]{Somerville2004, Bundy2005} and galaxies at very high redshift \citep[e.g.][]{Bouwens2006, Bouwens2008, Oesch2012}.  These surveys have led the way in
probing galaxy evolution from the earliest times up until the present
day, but they are fundamentally limited by the fact that they cannot
probe scales larger than a few Mpc, severely limiting investigations
of the environmental dependence of galaxy formation and evolution.

The goal of VIDEO is to perform an in-depth study of the
Universe over the redshift range $1<z<4$, linking the shallower
surveys such as the UKIDSS-LAS \citep[][]{Lawrence2007}, the VISTA Hemisphere Survey (VHS) and the VISTA Kilo-Degree Survey (VIKING) with the deeper UltraVISTA survey \citep{McCracken2012}. The depths
of the VIDEO survey have been chosen to detect galaxies at the knee in the galaxy luminosity function, $L^{\star}$ at
$z\sim\,4$, and down to $\sim\,0.1~L^{\star}$ at $z=1$, thereby enabling us to detect the bulk of the luminosity density arising from galaxies over 90 per cent of the history of the Universe and 
the most massive galaxies at the highest redshifts.  Thus we will be able to investigate in
exquisite detail which galaxies are in place first, and address
the issue of downsizing in the mass function of forming galaxies, where 
the massive early type galaxies appear to be in place before the less
massive galaxies \citep[e.g.][]{Cimatti2006, Fontanot2009}.  It is
also important to note that the epoch over which VIDEO is aimed is a
crucial one in the history of the Universe, as this is when the bulk
of the star formation \citep[e.g.][]{Madau1996,HopkinsBeacom2006,Lapi2011} and accretion  \citep[e.g.][]{Ueda2003, Croom2009} activity took place.  This is thus the ideal survey with which to investigate the effects that
star formation and accretion activity have on galaxy evolution in
general. Moreover, the intrinsic rarity of the most luminous AGN,
starburst and elliptical galaxies means that it is important to survey
a large enough area from which the luminosity function and clustering
of particular galaxy populations can be constrained, thereby enabling the evolution
of bias to be determined, improving on previous area-limited surveys that have measured the clustering of various galaxy populations to only $<0.5$\,degree scales \citep[e.g.][]{Foucaud2007, Coil2008,Hartley2010}.  Crucially, VIDEO will have sufficient area
to carry out these investigations as a function of both redshift and
environment.  

VIDEO will not only be able to detect the galaxies which
contribute the bulk of the luminosity density at these redshifts, but 
the five near-infrared filters, along with ancillary data from optical surveys and the {\em Spitzer Space Telescope}, will provide photometric redshifts for galaxies as faint as $L^{\star}$  all the way out to $z\sim\,4$, and to $z\sim\,6$ for the most massive galaxies.

\subsection{Statistical studies of the most massive galaxies at the highest redshifts}

In addition to probing the sub-$L^{\star}$ population up to $z \sim
4$, VIDEO will be uniquely placed to quantify the density and spatial
clustering of the most massive galaxies at the earliest epochs  \citep[e.g.][]{McLure2009,Hickey2010}. Initial work on this topic by \cite{Willott2012} used the combined data from the Canada-France-Hawaii Telescope Legacy Survey \citep[CFHTLS-D1; ][]{Ilbert2006}, along with the WIRCam Infrared Deep Survey \citep[WIRDS; ][]{Bielby2011}, UltraVISTA\citep{McCracken2012} and VIDEO to show that the bright end of the galaxy luminosity function at $z\sim 6$ follows an exponential decline and that these massive galaxies appear to be both larger and more dusty than their fainter counterparts discovered in deep HST imaging \citep[e.g.][]{Bouwens2006}.

With the full VIDEO survey area we will be able carry out the first statistically-significant
clustering analysis of massive galaxies towards the epoch of reionization, providing
a direct link between the underlying dark matter distribution and
galaxy populations and how this evolves up to the highest
redshifts. We note that the VIDEO survey will detect a factor of $\sim 10$
more $M>10^{11}$M$_{\odot}$ galaxies than UltraVISTA at $z \sim 6$,
as the limiting factor is area rather than depth. However, we emphasise that the combination of
VIDEO and UltraVISTA will provide the premier data set for
investigating the clustering properties of the highest redshift ($z >
4$) galaxies as a function of mass.



\subsection{Tracing the evolution of clusters from the formation epoch
  until the present day}

VIDEO provides data over the area and depth with which to study
the evolution of galaxy clusters from their formation epoch to the
present day.  Galaxy clusters are essential tracers of cosmic
evolution in the universe for two important reasons. First, clusters
are the largest virialized objects whose masses we can measure. 
Consequently,
comparisons of the present-day cluster mass distribution with the
distributions at earlier epochs can be used to determine the rate of
structure formation, placing constraints on cosmological
models \citep[see e.g. ][for a review]{Allen2011}. Second, the deep potential wells of clusters also mean that
they act as closed astrophysical laboratories that retain their
gaseous matter. Therefore clusters possess a wealth of information
about the processes associated with galaxy formation such as the
efficiency with which baryons are converted into stars and the effects
which feedback processes have on galaxy formation.

Between $0 < z <1$, clusters of galaxies appear to undergo little in the
way of strong evolution, either in their gas phase properties or in
the properties of the more massive galaxies within them \citep[e.g.][]{Rosati2002,Giodini2009, Collins2009}. The epoch at $z > 1$ is therefore a crucial one for studying
their evolution, which must be dramatic in the 4~Gyr between $1 < z <
4$ \citep[e.g.][]{vanBreukelen2006,Bielby2010}. The depth of VIDEO is such that it can
trace the bright end of the cluster luminosity function to $z=3$ and
look in detail at the less luminous cluster galaxies to $z=2$, while
its area should provide a sample of several tens of clusters with
masses above $10^{14}$~M$_{\odot}$ at $z>1$. Once found, such galaxy clusters can be studied in great detail in terms of their galaxy populations \citep[e.g.][]{Papovich2010,Tran2010}, and AGN content \citep[e.g.][]{vanBreukelen2009, Martini2009}, to trace how galaxies form and evolve in {\it closed} and dense environments. Furthermore, the shear number of clusters will allow the average properties of clusters to be studied in detail as a function of cosmic epoch \citep[][]{Geach2012}.

\subsection{Accretion activity over the history of the Universe}

\subsubsection{The first accreting black holes}

The SDSS has revolutionised studies of quasi-stellar objects (QSOs) at the highest redshifts \citep[e.g.][]{Fan2006},
providing the first evidence that the epoch of reionization was
coming to an end around $z \gtsim 6$ \citep{Becker2001, Fan2002}. Deeper optical surveys have since been pushing further down the QSO luminosity function \citep[e.g.][]{Willott2010}, but pushing to higher redshifts is
impossible with optical surveys, regardless of depth, due to the fact
that the Gunn-Peterson trough occupies all optical bands at $z >
6.5$. 


This work is being driven to higher redshift with near-infrared surveys such as the UKIDSS Large Area Survey, which recently discovered a QSO at $z = 7.085$ \citep{Mortlock2011} and the next step for this work will be with 
the wide-area VISTA public surveys such as VIKING and the VISTA
hemisphere survey, which will probe the bright end of the QSO luminosity
function at $z > 6$ \citep[e.g.][]{Findlay2012}. However, the shape of the QSO luminosity
function at these redshifts can only be studied with much deeper
near-infrared imaging over a significant survey area. This is the only
direct way to determine the contribution of accreting black holes to
the reionization of the Universe and constrain the density of
black-holes within the first Gyr after the Big Bang, as just using the
brightest quasars requires a huge extrapolation to fainter magnitudes.

Using the current best constraints on the QSO luminosity function \citep[][]{Willott2010} we expect 1-2 QSOs at $6.5<z<7.5$ in the full VIDEO survey; we will therefore be able to provide constraints on the faint end of the QSO luminosity function at $z>6$, a factor of 10 improvement in the number density of QSOs over narrower and deeper surveys such as UltraVISTA and the UKIDSS-UDS.

\subsubsection{The peak of accretion activity in the Universe}

With the VIDEO survey data we can also find lower-redshift QSOs which are
faint and/or reddened. However, interesting constraints could also be
made on the obscured AGN population when the VIDEO survey data are
combined with {\em Spitzer} \citep[e.g. SWIRE and SERVS;][]{Lonsdale2004, Mauduit2012}, X-ray \citep[e.g. XMM-LSS; ][]{Pierre2004} and {\em Herschel} \citep[e.g.][]{Hermes} observations, utilising near- and mid-infrared selection of AGN \citep[e.g.][]{Lacy2004,Stern2005,Maddox2008,Maddox2012,Banerji2012}. The depth
of VIDEO is crucial in determining the nature of distant obscured
AGN. Although the central quasar is obscured in these objects, VIDEO
will have the sensitivity to detect the host galaxies of such
sources up to $z \sim 5$ assuming that their host galaxies are $\sim
2-3~L^{\star}$ \citep[e.g.][]{Jarvis2001, Dunlop2003, Willott2003, Herbert2011}. 
The 12~degree$^2$ covered by VIDEO will ensure that the rarest and most luminous AGN are found and investigated in terms of their host galaxy properties and comoving space density. For example,  adopting the study of heavily obscured AGN from \cite{MartinezSansigre2005} we expect to find $\sim 300$ heavily obscured QSOs at $z>2$ in the VIDEO Survey. This would then allow the study of the range of host galaxy masses and the environmental density of these objects, something that would be severely restricted with COSMOS and UltraVISTA, and almost impossible with CANDELS,  due to the rarity of these luminous AGN.
Therefore, VIDEO is uniquely placed to probe the cosmic evolution of all AGN, both obscured and
unobscured and address the question of downsizing in the AGN population \citep[e.g.][]{Babic2007}.

\subsection{Supernovae}

There are several outstanding challenges to progress in Type Ia
Supernova (SN Ia) cosmology \citep{Conley2007,kessler2009,Conley2011}. The most
important include: i) Improved precision in the photometric
calibration, ii) Understanding SN Ia physics, particularly the SN
“colour laws” and the role of the progenitor system, iii) Constructing
comprehensive low-redshift surveys to anchor cosmological analyses,
and iv) Building the first homogeneous samples of SN~Ia at $z>1$.

Near-infrared observations are an essential tool to exploring these
trends further \citep[e.g.][]{Freedman2009}. As well as verifying the
extinction corrections made in the optical, theoretical SN Ia models
suggest that the intrinsic peak brightness dispersion could be substantially
smaller at longer wavelengths \citep{Kasen2006}. Observational studies
using $J$, $H$ photometry at $z<0.03$ show some evidence for this,
indicating that SNe Ia are equally good standard candles at these
wavelengths (peak dispersion
$<0.15$\,mag) and, have no apparent correlation with optical light
curve post-peak decline rate \citep{Wood-Vasey2008,BaroneNugent2012}.
Indeed, the combination of optical and near-IR data could provide the
strongest constraints on the optical extinction, enabling a
substantial improvement in SN Ia extinction corrections at all
wavelengths.

Further, as outlined by \cite{Kasen2006}, near-IR light curves offer
insight into several key aspects of SN Ia physics. The distinctive
light curve morphology is sensitive to both the amount and radial
distribution of iron group elements (radioactive and stable)
synthesised in the explosion. These in turn reflect the nature of the
explosion process and the progenitor star system, including the white
dwarf metallicity. Studies at near-IR wavelengths thus offer an
exciting opportunity to further understand and improve upon our dust
corrections of SNe Ia and the near-infrared light curves.

The VIDEO fields are key fields within the Dark Energy Survey (DES\footnote{http://www.darkenergysurvey.org})
supernova survey \citep{Bernstein2012} and will be able to
provide near-infrared data points on the light curves of supernovae
discovered in the DES Supernova survey. Furthermore, the spectral
energy distribution fitting and photometric redshift accuracy of the
supernovae host galaxies, which now play a key role in the
cosmological analyses \citep[][]{Sullivan2011}, will improve
substantially when the optical DES data are used in conjunction with
the deep VIDEO near-infrared data, and aid in the determination
of host galaxy stellar masses
\citep[][]{Kelly2010,Sullivan2010,Lampeitl2010}.

\section{Survey fields}

The VIDEO survey is being carried out over three of the most widely observed high-Galactic-latitude fields, these are two VISTA tiles ($\sim 3$~degree$^2$) in the ELAIS-S1 field, three VISTA ($\sim 4.5$~degree$^2$) tiles in the XMM-Newton Large Scale Structure field, and another three tiles in the extended Chandra Deep Field South, covering a total area of $\sim 12$~degree$^2$. These fields have a wealth of data from the X-rays through to the radio waveband, and are, along with COSMOS/UltraVISTA, the primary fields for observations with future facilities in the southern hemisphere. 

\begin{figure}
\includegraphics[width=0.45\textwidth]{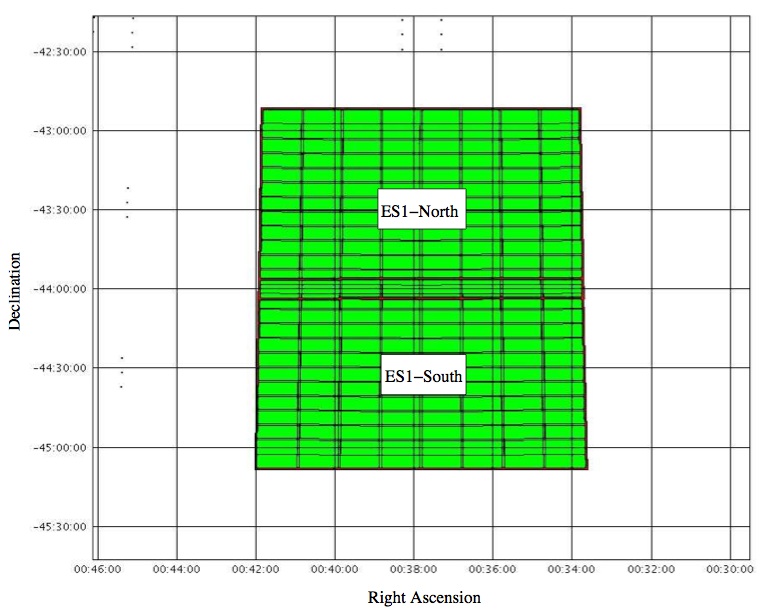}
\includegraphics[width=0.45\textwidth]{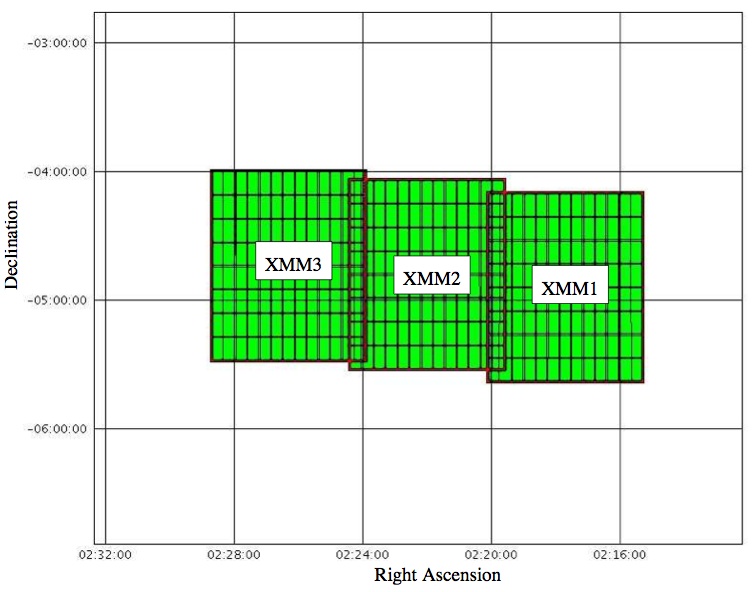}
\includegraphics[width=0.45\textwidth]{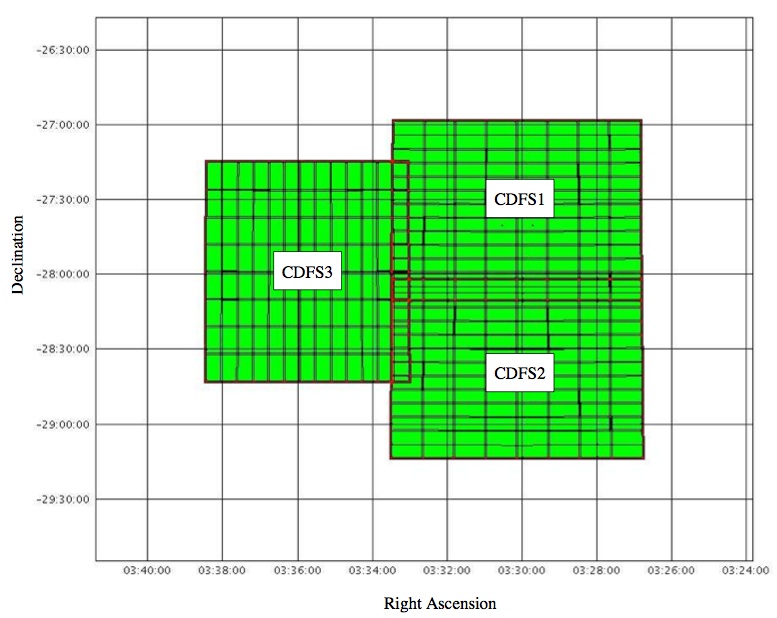}
\caption{The three areas of sky over which the VIDEO survey will be carried out. From top to bottom; ELAIS-S1 (3~degree$^2$), XMM-LSS (4.5~degree$^2$) and the Extended Chandra Deep Field South (4.5~degree$^2$). }
\label{fig:fields}
\end{figure}

Figure~\ref{fig:fields} shows the footprint of the VIDEO observations in each of these three fields. One can see that the individual tiles are distributed in different ways. The design of these positions was driven by the need to cover the wealth of ancillary data available in these fields, including {\em Spitzer} \citep{Lonsdale2004,Mauduit2012} and  {\em Herschel} \citep{Hermes}, in addition to future ground-based optical  data from both the VLT Survey Telescope and the Dark Energy Survey. Further in the future these areas will also be targeted with the deepest radio observations with the Square Kilometre Array precursor telescope MeerKAT \citep{Jonas2009} as part of the deep continuum survey \citep{Jarvis2012}.

\section{Survey depth}

Our aim is to be able to detect a galaxy at or below the break of the galaxy luminosity function over 90 per cent of the age of the Universe, which
corresponds to $0 < z< 4$. This ensures that we are sensitive to the bulk of the energy density from galaxies at one
of the most important epochs in the Universe, where star formation and
accretion activity were at a maximum. 

We use the $K-$band luminosity function from Cirasuolo et al. al.\ 2010) where $M^{\star} = -22.26$ at $z=0$. Assuming a passively evolving
stellar population with a high formation redshift, then we expect a $z \sim 4$ galaxy to have a total magnitude of $K \sim 23.2$.  Given the seeing constraints (see section~\ref{sec:observations}) and accounting for the apparent size evolution of massive passive galaxies at high redshift \citep[e.g.][]{Zirm2007,vanDokkum2010, McLure2012} we therefore define our 5$\sigma$ $K_{\rm s}$-band limit as $K_{\rm s} = 23.5$ in a 2~arcsec diameter aperture.

The survey limits in the shorter near-infrared filters ($ZJH$) are then determined assuming an elliptical galaxy template at $z\sim\,4$. The $Y-$band observations are slightly shallower, and are fixed to that of a flat spectral energy distribution from the $J-$band to help improve photometric redshift accuracy at redshifts where the 4000\AA\, Balmer and the Lyman-$\alpha$ breaks are redshifted between the $Z-$ and $J-$band filters.

\begin{table*}
\caption{Observational strategy for the VIDEO survey, including the planned integration times and sky conditions for the various VIDEO filters. We note that prior to Period 86 the $Z-$band observations were taken with DIT$\times$NDIT=$45\times 1$~seconds and Jitters$\times$ N$_{\rm obs}=10\times 61$. The $5\sigma$ limits are for a point source in a 2\,arcsec diameter aperture and are the expected depths prior to the survey starting.}\label{tab:observations}
\begin{tabular}{|ccccccccc|}
\hline
Filter & Planned Time (h) & DIT$\times$NDIT & Jitters$\times$N$_{\rm obs}$ & 5$\sigma$ & $2^{\prime\prime}$ ap.mag. & Seeing & Moon   \\
    & (per source) & (seconds) && AB & Vega &  (arcsec) & \\ 
\hline
Z & 17.0 & $50\times 1$ & $9\times 68$  & 25.7 & 25.2    & 0.8 & Dark \\
Y & 7.5  & $30 \times 2$& $8\times 28$ & 24.6 & 24.0    & 0.8 & Grey \\ 
J & 9.6  & $30\times 2$ & $8\times 36$  & 24.5 & 23.7  & 0.8 & Grey \\ 
H & 9.1  & $10\times 6$ & $7\times 39$& 24.0 & 22.7 & 0.8 & Bright \\ 
K$_{s}$   & 9.8 & $10\times 6$ & $7\times 42$ & 23.5 & 21.7 & 0.8 & Bright  \\
\hline
\end{tabular}
\end{table*}

\section{Observations}\label{sec:observations}

Here we describe the properties of the VIDEO survey data taken between 2009-11-03 and 2011-11-20 over the VIDEO-XMM3 field (see Fig.~\ref{fig:fields}). We describe the observations for this single tile, noting that the observational strategy for the other seven tiles within the full VIDEO survey will be the same. We note that the VIDEO survey is an ongoing project and that the final depth of the $Y-$band data  in this field will be deeper than presented here by $\sim 0.1$~mag.

The VISTA Infrared Camera \citep[VIRCAM; ][]{Dalton2006} is a wide-field near-infrared camera consisting of sixteen $2048 \times 2048$ Raytheon VIRGO HgCdTe arrays distributed over the focal plane of VISTA. The detectors present  a sparsely-filled focal plane with gaps between each array of 0.90 and 0.425 of a detector along the  $x$ and $y$ axes respectively \citep{Emerson2004}. The mean pixel scale is 0.34$^{\prime\prime}$\,pixel$^{-1}$.

\begin{figure}
\includegraphics[width=0.98\columnwidth]{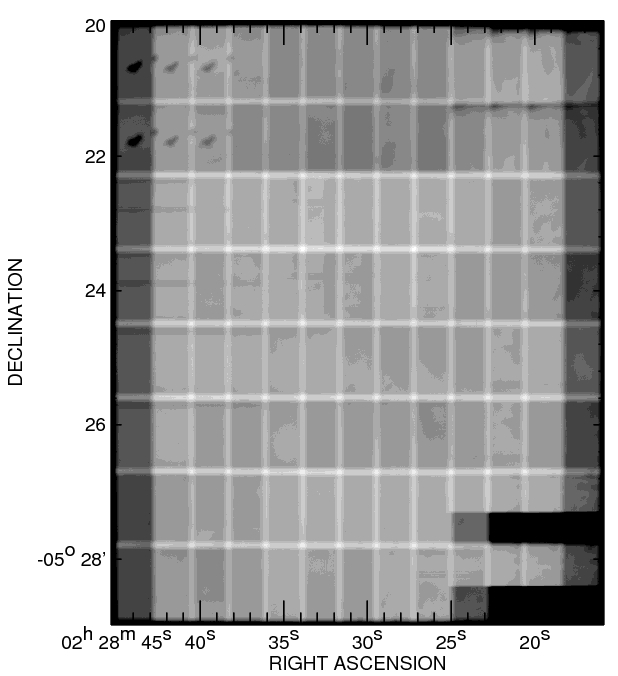}
\caption{Confidence map of the VIDEO-XMM3 image from the VIDEO $K_{\rm s}-$band. The dark regions correspond to areas of low confidence due to lower exposure times through the jittering and/or due to detector artefacts and edges. In particular, a region of dead pixels is evident in the upper left detected (detector 1) and a region of increased noise is present in the lower right region in detector 16 (see text for details).  
The regions of overlap between the individual detectors also have more than the nominal exposure time.
Note that this and some other fields are rotated by 90$^{\circ}$ with respect to the nominal camera rotation.}
\label{fig:confidence}
\end{figure}

As the sky is both bright (particularly in the $H$ and $K_{\rm s}$-band) and highly variable (due to OH lines in the atmosphere) at near-infrared wavelengths it is necessary to split the individual exposures into a series of short exposures and then add these together. To overcome the issue of the sky background saturating the detectors, particularly at the longer wavelengths, the observations are split into a number of NDIT times the Detector Integration Times (DITs). 
Thus for a single pointing the total time spent in a single position on the sky
is given by DIT $\times$ NDIT. As can be seen in Table~\ref{tab:observations}, the DITs be as short as 10~seconds for the $H-$ and $K_{\rm s}-$bands. To overcome the variable sky fluctuations and bad pixels, the telescope is  ``jittered'' around by a random offset of $\ltsim\,20$~arcsec in right ascension and declination and the series of DIT$\times$ NDIT is repeated at each jittered position.  This also ensures that the same pattern is not repeated throughout the survey period which would result in bad pixels and detector artefacts adding together after combining the full survey data, although we note that some very large detector defects are impossible to remove (Fig.~\ref{fig:confidence}).


The paw-print of the 16 detectors provides a non-contiguous instantaneous field-of-view of $\sim 0.6$~degree$^{2}$. A full tile, where the gaps between detectors are filled, requires moving the telescope in a sequence of six offsets in both right ascension and declination, which together produce a contiguous field-of-view of
$\sim 1 \times 1.5$~degree$^2$ which (for no jitters) would be sampled at least twice, with some positions being sampled up to six times. In addition there are singly sampled strips of width $\sim 0.1$~degrees at the extremities of the longer dimensions of the tiles. These characteristics can be seen in the 
confidence map of the complete VIDEO-XMM3 tile in the
$K_{\rm s}−$band which is presented in Fig.~\ref{fig:confidence} and shows the regions of overlap
between the individual detectors which have more than the nominal
exposure time, and also darker, underexposed $\sim 0.1$~degree vertical strips at the edges, and other dark areas due to detector artefacts. The prominent detector artefact regions at the top left result from islands of bad pixels on detector 1, and the areas in the lower right result from the half of detector 16 which is affected by a time varying quantum efficiency (most notably in the shorter wavelength filters) which makes flat fielding that detector extremely challenging.

The number of jitters and paw-prints that are carried out at each pointing is constrained by the limited time per observation block (OB) of 1\,hour and 5\,min, set by ESO to ensure flexible observing over the course of the night and to mitigate changes in the weather conditions over the course of an OB. 

All VIDEO survey observations are carried out in good seeing conditions, with the seeing in each filter required to have FWHM $<0.8$\arcsec, ensuring very good data quality and that the full depth is achieved in the estimated time, although in practice observations with 10 per cent worse seeing are formally accepted (i.e. up to 0.9\arcsec) as the seeing can change during a single OB.

Full details of the DIT, NDIT and number of jitters are provided in Table~\ref{tab:observations} along with the total number of hours required to complete a tile, the expected depth and the Moon phase at which each filter is observed.

\section{Data Reduction}\label{sec:reduction}

\begin{figure*}
\includegraphics[width=0.94\textwidth, angle=0]{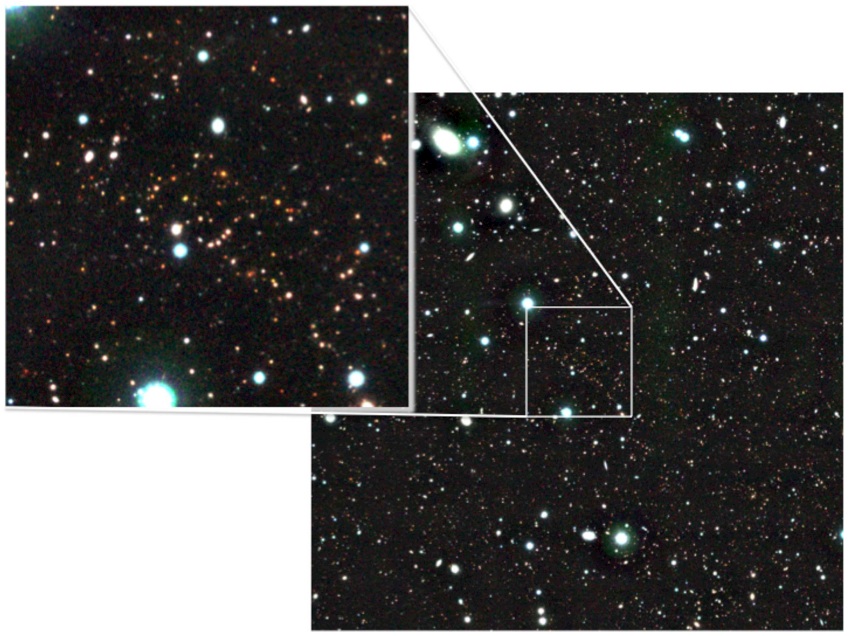}
\caption{Example $YJK_{\rm s}$ colour image of a $15\times15$\,arcmin$^2$ region within the VIDEO-XMM3 field centred on the putative $z\sim 2$ galaxy cluster JKCS041 \citep{Andreon2011}. The zoomed-in region has a size of $3 \times 3$~arcmin$^{2}$ showing the depth and image resolution of the VIDEO data. }
\label{fig:3col_image}
\end{figure*}

Initial data reduction steps are performed at the Cambridge Astronomical Survey Unit (CASU) 
using a software pipeline developed specifically for the reduction of VIRCAM data, as part of the VISTA Data Flow System (VDFS) described by \cite{Irwin2004} and updated on the CASU webpages\footnote{http://casu.ast.cam.ac.uk/surveys-projects/vista/technical/data-processing}. This pipeline is modular, and allows different processing recipes to be applied to data obtained with different observing strategies.

For the VIDEO survey, the following steps are applied to each raw data frame (itself the result of NDIT double-correlated-sampling exposures which are co-added by the data acquisition system):
\begin{itemize}

\item {\em Destriping} -- removes a low-level horizontal stripe pattern introduced by the controller and correlates over the four detectors on each controller,

\item  {\em linearity correction} --  corrects for the non-linear detector response which is typically 2-4 per cent depending on the detector,

\item {\em  dark correction} -- subtracts a mean dark exposure from the image, correcting for dark current and some other electronic effects,

\item {\em flat field correction} -- divides images by a mean twilight sky image, to correct for position-dependent variations in telescope/camera throughput and detector response,

\item {\em sky background correction} -- subtracts an estimate of the atmospheric emission (more details below),

\item {\em astrometric calibration} -- compares the positions
of stars in the image with those in the 2MASS
point source catalogue \citep{2mass},

\item {\em photometric calibration} -- calculates magnitudes in the VISTA Vega-magnitude photometric system\footnote{http://casu.ast.cam.ac.uk/surveys-projects/vista/technical/data-processing} for unsaturated 2MASS stars in the image using their magnitudes and colours in the 2MASS
point source catalogue, to set the zero point in the VISTA photometric system, and

\item {\em jitter stacking} -- combines the jittered images in a single paw-print position, using a mean stack of bi-linearly-resampled images, with outlier rejection to remove cosmic rays, fast-moving objects, and bad pixels.
\end{itemize}

Most of these processing steps are identical to those described by \cite{Irwin2004} and applied to other VISTA survey data, but the sky background estimation has been tuned for the needs of VIDEO. The faint fluxes probed by VIDEO make it particularly sensitive to the sky subtraction strategy, because many of the most interesting sources in the VIDEO images are indistinguishable from noise until many (sky-subtracted) paw-prints are stacked, and are thus difficult to eliminate from estimates of the background (which must necessarily be based on images taken close to the image from which the background is to be subtracted).

The strategy used for sky estimation in the current VIDEO release is to combine, with outlier rejection, the jittered images taken in three paw-prints of a tile (which for VIDEO means they are obtained from $20-29$ individual paw-prints, over a period of $\sim30$ minutes from either the first or the second half of the OB) without shifting them to match astrometry. Bright objects detected in these images are masked out, but objects below the noise level are not. To prevent below-noise objects contributing to the sky estimate (and causing sky subtraction to reduce their fluxes systematically), the image which is having its background subtracted is omitted from the background estimate. The sky image is then scaled and subtracted from each individual paw-print.

The full-depth image stacks have been co-added using {\sc SWarp}\footnote{http://www.astromatic.net/software/swarp} and are weighted mean stacks of the paw-print stacks produced by CASU, after rejecting the paw-print stacks with seeing worse than 0.9\arcsec\, FWHM.  We turn on the background subtraction within SWarp (estimated using a mesh size of 128 pixels) to remove large-scale gradients which remain after the sky-subtraction step. We note that using SWarp introduces correlated pixel noise and therefore simple noise measurements to estimate flux errors and sensitivity limits are not reliable (see Section~\ref{sec:noise}).
A $15\times 15$~arcmin$^{2}$ $YJK_{\rm s}$ image from within the VIDEO field is shown in Fig.~\ref{fig:3col_image}, along with a zoomed in image of the central $3\times 3$~arcmin$^2$ region, demonstrating the quality of the data.

Catalogues have been extracted from each set of images using multiple passes of {\sc SExtractor} \citep{sextractor} in double-image mode. Each tile catalogue contains objects detected in any of the $K_{\rm s}$, $H$, $J$, $Y$, or $Z$ bands, with measurements made in all the other bands based on the position in the detection image. Duplicate detections of objects are removed, by retaining only the longest wavelength detection (after matching the $K_{\rm s}-$, $H-$, $J-$, $Y-$ and $Z-$band detected catalogues with a 1 arcsecond tolerance). 
We extract photometry in fixed circular apertures 1, 2, 3, 4, and 5 arcsecs in diameter, plus Kron and Petrosian fluxes. To correct for correlated noise in the images, the errors for fixed aperture photometry are estimated by measuring the RMS flux in randomly placed apertures of the same size, then adding Poisson errors based on object counts. Errors for Kron and Petrosian fluxes are corrected by scaling the random (non-Poisson) component of the error (as estimated by {\sc SExtractor}); the scaling factor is estimated by comparing the median {\sc SExtractor} error for randomly placed 5~arcsec apertures with the actual standard deviation of fluxes in those apertures.

Bright stars are not masked out from the images before catalogue extraction. Ghosting, especially in the $Y-$ and $Z-$bands, causes large halos around the brightest stars. A region file was constructed to enclose the regions visibly affected by these ghosts, and all objects whose centres fall within these regions have {\sc HALO\_FLAG} = 1 in the catalogue.

These catalogues are intended to be complementary to those which will be hosted by the ESO and VISTA Science Archive \citep[VSA; ][]{Cross2012}. 

\section{Final Stacked Images}\label{sec:finalstacks}

\subsection{Noise measurements}\label{sec:noise}
The individual VIDEO tiles are produced by co-adding a series of jittered and shifted VISTA paw-prints, where every pixel in the detectors has been exposed on at least two different paw prints. We measure the depth of these individual tiles by placing apertures on parts of the images at least 10 arcsec away from any detected objects in order to measure the level of the sky background in a range of apertures. We then fit two half Gaussians to the positive and negative noise distribution and take the standard deviation ($\sigma$) from the negative side of the Gaussian as the noise estimate. We only use the negative part of the Gaussian fit as this is uncontaminated by real sources in the image that may contribute to and overestimate the noise on the positive side, although we note that in practice this has little effect. Determining the noise by this method takes into account the correlated noise in the deep stacked images and provides a realistic estimate that includes any flat-field fluctuations between observations.
The noise properties of the final tiled image are well represented by a Gaussian distribution. Table~\ref{tab:realdepth} provides the measured background noise of the VIDEO XMM-3 tile in all filters as a function of aperture diameter. The total integration time relating to the noise measurement is also shown.


\begin{table}
\caption{Measured 5$\sigma$ depths for the VIDEO-XMM3 field in $1-5$\,arcsec diameter apertures in AB magnitudes. The time given is the total integration time of the images used to measure these. }\label{tab:realdepth}
\begin{tabular}{|cccccccccc|}
\hline
Filter & Time (h)  & $1^{\prime\prime}$  & $2^{\prime\prime}$. & $3^{\prime\prime}$ & $4^{\prime\prime}$ & $5^{\prime\prime}$\\
&   (per pixel) & (5$\sigma$) & (5$\sigma$) & (5$\sigma$) & (5$\sigma$) & (5$\sigma$)  \\ 
\hline
Z &17.13 & 26.89 & 25.66 & 24.87 & 24.33 & 23.83\\
Y & 6.05 & 25.60 & 24.51 & 23.74 & 23.18 & 22.73 \\
J & 9.65 & 25.59 & 24.44 & 23.70 & 23.11 & 22.63 \\
H & 8.13 & 25.25 & 24.12 & 23.42 & 22.87 & 22.40 \\
K$_{s}$ & 9.03  & 24.86 & 23.77 & 23.10 & 22.53 & 22.13 \\
\hline
\end{tabular}
\end{table}


\subsection{Completeness and Reliability}

\begin{figure}
\includegraphics[width=0.98\columnwidth]{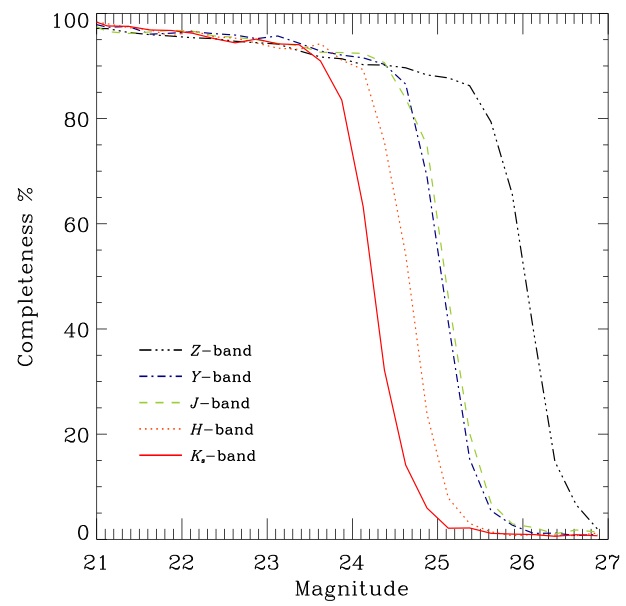}
\caption{Completeness as a function of magnitude measured in a 2~arcsec diameter aperture for all five VIDEO filters. Note that the $Y-$band data are not at their final depth (see text for details).}
\label{fig:completeness}
\end{figure}

Our ability to recover objects as a function of the objects' flux is dependent on both the noise properties of the images and the proximity to nearby bright objects and other objects in the field. In order to measure the completeness of the deep VIDEO stacked data we add point sources with a range of magnitudes spanning $15<K_{\rm s}<25$ with FWHM consistent with the average seeing measurement across the image and re-extract the  source catalogues using the same {\sc SExtractor} parameters as for the real sources. We then compare the input and extracted catalogues to determine the completeness of each stacked tile. Fig.~\ref{fig:completeness} shows the completeness of the VIDEO survey for all filters and Table~\ref{tab:completeness} provides the values of the completeness down to the $5\sigma$ limit in each filter. We find that the data are typically complete at the 80--90 per cent level for the $5\sigma$ depths given in Table~\ref{tab:realdepth}.  We note that the completeness for extended sources will be worse than the completeness estimates for point sources presented here.

We are also able to determine the reliability of the source extraction multiplying the stacked image by $-1$ and re-running {\sc SExtractor}  over this image. The number of extracted sources, thus provides a good estimate of the number of spurious sources in the extracted catalogues. Down to the $5\sigma$ flux-density limit of $K_{\rm s} \le\,23.77$ we find that spurious sources contribute at the $\sim 0.07$~per cent level.

\begin{table}
\caption{Completeness corrections for a point source in the five VIDEO filters in the VIDEO-XMM3 field. The completeness corrections are provided up to and including the $5\sigma$ depth for a 2~arcsec diameter aperture in the respective filters.}\label{tab:completeness}
\begin{tabular}{cccccc}
\hline
Magnitude& \multicolumn{5}{c}{ Completeness } \\
 &   $Z$  & $Y$  &  $J$  & $H$ & $K_{\rm s}$ \\
\hline
17.25 & 0.996 &  0.999 & 0.999&  0.999 &0.998\\
17.75 & 0.995 &  0.998 & 0.999&  0.995 &0.996\\
18.25 & 0.994 &  0.996 & 0.998&  0.999 &0.997\\
18.75 & 0.993 &  0.995 & 0.994&  0.996 &0.994\\
19.25 & 0.990 &  0.994 & 0.993&  0.994 &0.995\\
19.75 & 0.988&  0.993 & 0.993&  0.995 &0.991\\
20.25 & 0.983 &  0.986 & 0.986&  0.983 &0.986\\
20.75 & 0.974 &  0.984 & 0.984&  0.987 &0.986\\
21.25 & 0.967 &  0.980 & 0.975&  0.985 &0.976\\
21.75 & 0.958 &  0.973 & 0.970&  0.969 &0.975\\
22.25 & 0.953 &  0.963 & 0.948&  0.967 &0.962\\
22.75 & 0.946 &  0.959 & 0.956&  0.951 &0.957\\
23.25 & 0.935 &  0.948 & 0.938&  0.937 &0.955\\
23.75 & 0.916 &  0.939 & 0.923&  0.918 &0.884\\
24.25 & 0.902 &  0.924 & 0.916&  0.846 &---\\
24.75 & 0.890 &  0.759 & 0.818&  ---&---\\
25.25 & 0.871 &  --- & --- & ---  & ---\\
25.75 & 0.721 &  --- & --- & ---  & ---\\
\hline
\end{tabular}
\end{table}

\subsection{Astrometric Comparison}

\begin{figure}
\includegraphics[width=0.48\textwidth]{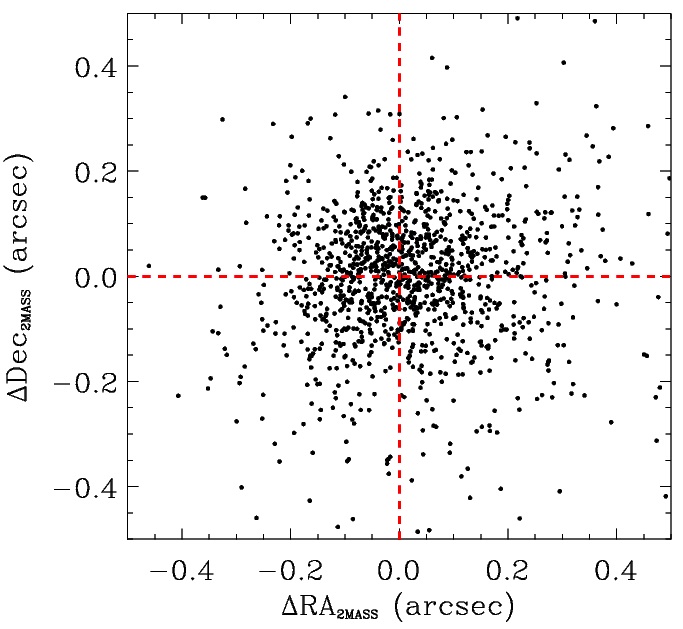}
\caption{Astrometric comparison between the VIDEO $K_{\rm s}-$band image and the astrometry from 2MASS sources in the VIDEO-XMM3 field. The red dashed lines show the offsets corresponding to $\Delta$ RA=0~arcsec and $\Delta$ Dec=0~arcsec}\label{fig:astrom2mass}
 \end{figure}

We compare our astrometric calibration over the VIDEO-XMM3 field with the Two-Micron All Sky Survey \citep[2MASS; ][]{2mass}. 2MASS is used to define the astrometric solution for VIDEO on a paw-print by paw-print basis, we therefore expect close agreement. Fig.~\ref{fig:astrom2mass} shows the distribution for the difference in Right Ascension and Declination for all of the objects detected with a signal-to-noise ratio $> 10$ in the 2MASS $K_{\rm s}$-band over the VIDEO field. One can immediately see that the distribution is strongly peaked around $\Delta$ RA $\sim 0$ and $\Delta$ Dec $\sim 0$, with mean offsets of $\alpha_{\rm VIDEO}-\alpha_{\rm 2MASS} = 0.020 \pm 0.006$~arcsec with an rms of 0.288~arcsec, and $\delta_{\rm VIDEO}-\delta_{\rm 2MASS} = -0.004 \pm 0.005$~arcsec with an rms of 0.253~arcsec. The median offsets are $\alpha_{\rm VIDEO}-\alpha_{\rm 2MASS} = 0.003 \pm 0.006$~arcsec and $\delta_{\rm VIDEO}-\delta_{\rm 2MASS} = 0.007 \pm 0.005$~arcsec. Thus, the VIDEO astrometry is entirely consistent with the 2MASS astrometric calibration with no significant offset.


The
excellent agreement between VIDEO and 2MASS is unsurprising given that
2MASS is used for the astrometric calibration of VIDEO,  but the small spread indicates that the data are
well-calibrated across the whole tile.

\subsection{Photometric Comparison}\label{sec:compare_phot}

\begin{figure}
\includegraphics[width=0.48\textwidth]{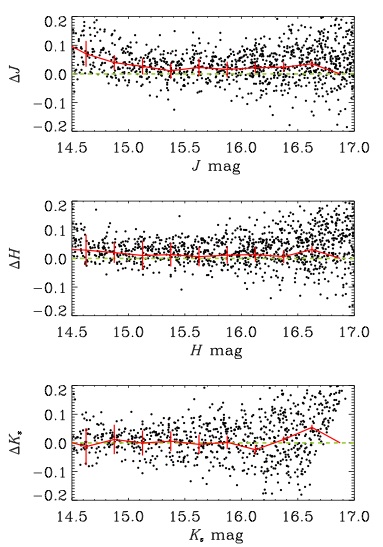}
\caption{Comparison of photometry between VIDEO and 2MASS in the $J$, $H$ and $K_{\rm s}$-band filters. $\Delta$~mag corresponds to the VIDEO photometry (corrected to the 2MASS magnitude system) minus the 2MASS photometry in each case. The dashed green line shows the one-to-one relation and the solid red line denotes the median difference as a function of magnitude.}
\label{fig:phot2mass}
\end{figure}

The VISTA photometric system is naturally not identical to any other photometric system and
transformations to other photometric systems are not yet available. Nevertheless it is interesting to use the 2MASS and WIRDS data sets to check the
photometric calibration of the VIDEO data set in the $J$, $H$ and
$K_{\rm s}$ filters. We again select all 2MASS objects with SNR $> 10$ that fall within the VIDEO-XMM3 tile and cross-match these with the VIDEO catalogue in each filter. As the resolution of 2MASS is much poorer than VIDEO we use the total magnitudes measurements from 2MASS and the {\sc MAG\_AUTO} measurements from the VIDEO catalogue which we consider to be the closest to total magnitudes for the VIDEO data set. Fig.~\ref{fig:phot2mass} shows the difference in magnitude for the sources detected in 2MASS, adjusting the magnitudes in the VIDEO bands according to the colour equations derived from 2MASS\footnote{http://casu.ast.cam.ac.uk/surveys-projects/vista/technical/vista-sensitivity}, following \cite{Hodgkin2009}. Given the derived colour equations we expect a close correspondence, and for the limited magnitude range where there is overlap, we do indeed find very good consistency across all magnitudes to which 2MASS is sensitive.

\begin{figure}
\includegraphics[width=0.48\textwidth]{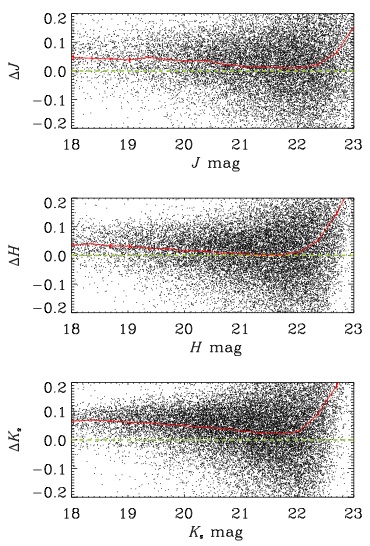}
\caption{Comparison of photometry between VIDEO and WIRDS in the $J$, $H$ and $K_{\rm s}$-band filters. $\Delta$~mag corresponds to the VIDEO photometry minus the WIRDS photometry in each case. The dashed green line shows the one-to-one relation and the solid red line denotes the median difference as a function of magnitude}
\label{fig:photwirds}
\end{figure}

\begin{figure}
\includegraphics[width=0.48\textwidth]{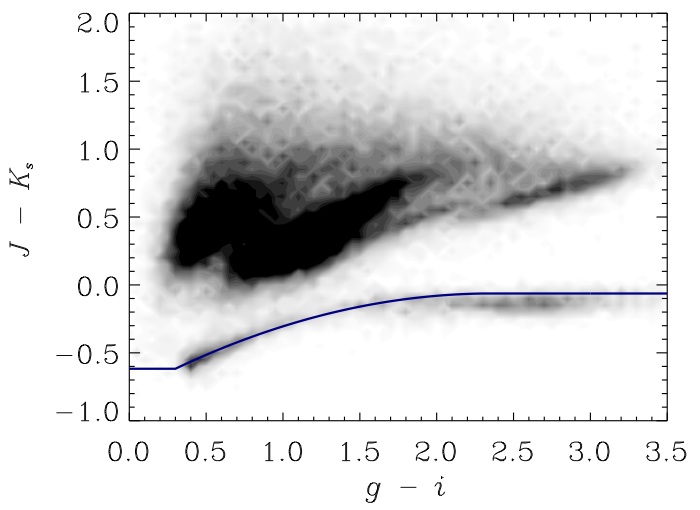}
\caption{Colour-colour diagram showing $g-i$ from the CFHTLS-D1 versus $J-K_{\rm s}$ for sources with $K_{\rm s} < 23.5$ in VIDEO. The solid blue line denotes the stellar locus fit from \citep{Baldry2010} offset by 0.1\,mag in $J-K_{\rm s}$ colour as described in the text.}
\label{fig:gi_jk}
\end{figure}

\begin{figure}
\includegraphics[width=0.48\textwidth]{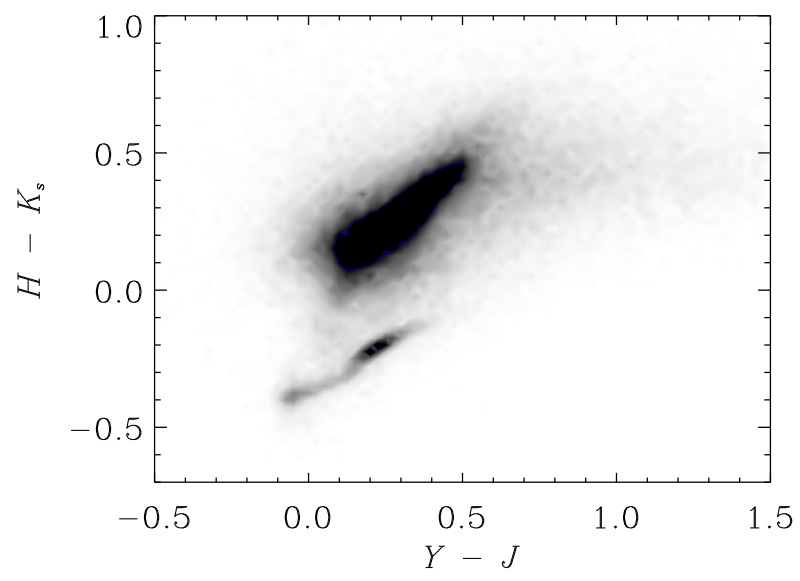}
\caption{Near-infrared colour-colour  diagram for sources with $K_{\rm s} < 23.5$ in VIDEO. The region populated by the majority of stars can easily be seen as the lower clump of objects. }
\label{fig:YJ_HK}
\end{figure}

Moving to much fainter magnitudes, we compare the photometry between the WIRDS $J$, $H$ and $K_{\rm s}$-band data and the VIDEO data in Fig.~\ref{fig:photwirds}. Again using the {\sc MAG\_AUTO} magnitude and correcting the VIDEO magnitudes to 2MASS magnitudes as detailed above and correcting the WIRDS magnitudes according to the values given in \cite{Bielby2011}, we find reasonable consistency between the two data sets in all three filters where data is available. However, there is a tendency for $\Delta$\,mag  to increase towards bright magnitudes. This is attributed to the lack of a well-defined colour term in converting from the WIRDS photometry to the 2MASS photometry.

\subsection{Number counts}\label{sec:numcounts}

\begin{figure*}
\includegraphics[width=0.33\textwidth]{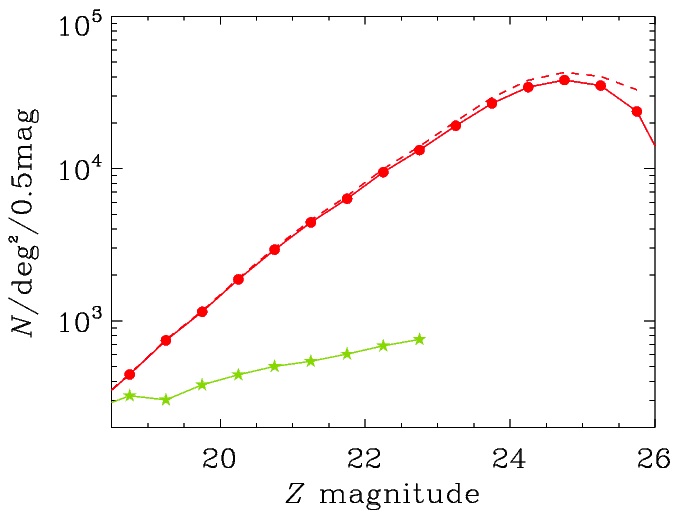}
\includegraphics[width=0.33\textwidth]{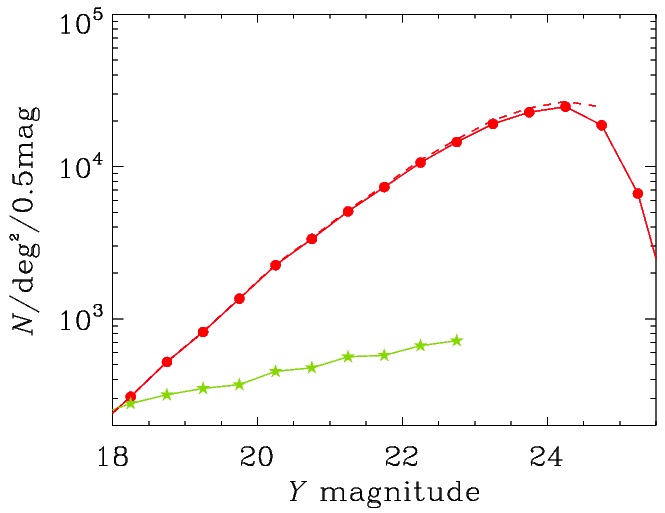}
\includegraphics[width=0.33\textwidth]{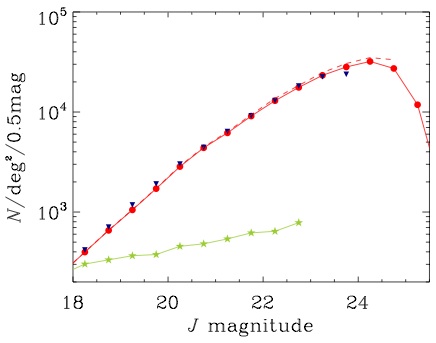}
\end{figure*}
\begin{figure*}
\includegraphics[width=0.33\textwidth]{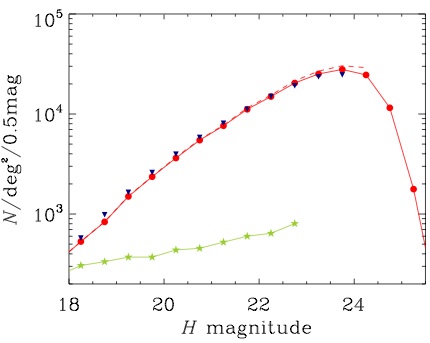}
\includegraphics[width=0.33\textwidth]{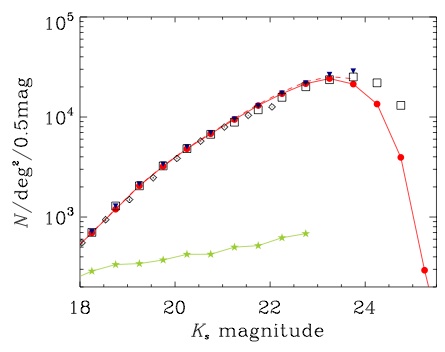}
\caption{Total magnitude differential number counts for galaxies detected in the $Z$, $Y$, $J$, $H$ and $K_{\rm s}$-bands in the VIDEO-XMM3 field. Filled red circles and solid red line show the number counts uncorrected for completeness and the red dashed line shows the completeness corrected source counts using the values from Table~\ref{tab:completeness}. The Poisson uncertainties are smaller than the symbol size. The stellar number counts, separated as discussed in section~\ref{sec:numcounts} are also shown as green stars. Also shown are the number counts from other studies in the filters available, these are; CFHT-WIRDS \citep[inverted filled blue triangles; ][]{Bielby2011}, the UKIDSS-DXS \citep[black diamonds; ][]{Kim2011} and the UltraVISTA \citep[open black squares; ][]{McCracken2012}, showing excellent agreement.}
\label{fig:ncounts}
\end{figure*}

For measuring the number counts in the VIDEO data we use the MAG\_AUTO magnitudes and the $g-i$ versus $J-K_{\rm s}$ colour-colour diagram to separate galaxies from stars. Following \citet{Baldry2010} we fit a quadratic function to the stellar locus and separate the galaxies from the stars using the distance from the locus of the $J-K_{\rm s}$ colour \citep[see equation~2 in ][]{Baldry2010}. For VIDEO we find that a slight adjustment of the location of locus defined by \citet{Baldry2010} along the $J-K$ axis is sufficient to account for the difference between the $K$ filter in UKIDSS and the $K_{\rm s}$ filter in VISTA. We therefore find
\begin{equation*}
\begin{array}{lrr}
f_{\rm locus} (x)= & -0.6127 & x \le 0.3\\
f_{\rm locus}(x)=& -0.79+0.615x-0.13x^2 & {\rm for~~~ } 0.3< x\le 2.3\\
f_{\rm locus}(x)=& -0.0632 & x>2.3\\
\end{array}
\end{equation*}
where $x = g-i$. The $g-i$ versus $J-K_{\rm s}$ colour-colour diagram is shown in Fig.~\ref{fig:gi_jk} along with the locus defined above. We then separate stars from galaxies using a divide at $J-K_{\rm s} = f_{\rm locus}+0.1$\,mag, where galaxies lie above this line and stars below it. We note that it is possible to identify the stellar locus in just the VIDEO filters (Fig.~\ref{fig:YJ_HK}), although this is less precise than the adopted method.

\begin{table*}
\caption{Differential total-magnitude galaxy counts per degree$^2$ per 0.5~mag bin for the five VIDEO filters, not corrected for completeness.}\label{tab:ncounts}
\begin{tabular}{crlrlrlrlrl}
\hline
& \multicolumn{10}{c}{  $N$ /\, degree$^{2}$/ 0.5\,mag} \\
Magnitude  &   \multicolumn{2}{c}{ $Z$}  & \multicolumn{2}{c}{ $Y$}  & \multicolumn{2}{c}{ $J$}  & \multicolumn{2}{c}{ $H$}  & \multicolumn{2}{c}{ $K_{\rm s}$}   \\
\hline
17.25 &   94& $\pm$     10&   119 &  $\pm$    11 &   158 & $\pm$    13 &   178 & $\pm$    14&   242& $\pm$    16 \\
17.75 &  173 & $\pm$    13 &   188&  $\pm$    14 &   240 & $\pm$    16 &   330 & $\pm$    18 &   408 & $\pm$    21 \\
18.25 &  276 & $\pm$    17 &   308 & $\pm$    18 &   398 & $\pm$    20 &   531 & $\pm$    23 &   695 & $\pm$    27 \\
18.75 &  445 & $\pm$    21 &   521 & $\pm$    23 &   655 & $\pm$    26 &   835 & $\pm$    29 &  1196 & $\pm$    35 \\
19.25 &  744 & $\pm$    28 &   820 & $\pm$    29 &  1050 & $\pm$    33 &  1499 & $\pm$    39 &  2054 & $\pm$    46 \\
19.75 & 1147 & $\pm$    34 &  1357 & $\pm$    37 &  1712 & $\pm$    42 &  2361 & $\pm$    49 &  3213 & $\pm$    58 \\
20.25 & 1873 & $\pm$    44 &  2253 & $\pm$    48 &  2846 & $\pm$    54 &  3630 & $\pm$    61 &  4809 & $\pm$    70 \\
20.75 & 2934 & $\pm$    55 &  3353 & $\pm$    59 &  4386 & $\pm$    67 &  5477 & $\pm$    75 &  6804 & $\pm$    84 \\
21.25 & 4435 & $\pm$    68 &  5079 & $\pm$    72 &  6184 & $\pm$    80 &  7625 & $\pm$    89 &  9436 & $\pm$    99 \\
21.75 & 6339 & $\pm$    81 &  7341 & $\pm$    87 &  9104 & $\pm$    97 & 11140 & $\pm$   107 & 13025 & $\pm$   116 \\
22.25 & 9483 & $\pm$    99 & 10631 & $\pm$   105 & 12961 & $\pm$   116 & 14943 & $\pm$   124 & 17100 & $\pm$   133 \\
22.75 &13244 & $\pm$   117 & 14521 & $\pm$   122 & 17609 & $\pm$   135 & 20408 & $\pm$   145 & 21525 & $\pm$   149 \\
23.25 &19134 & $\pm$   140 & 19110 & $\pm$   140 & 23267 & $\pm$   155 & 25138 & $\pm$   161 & 24232 & $\pm$   158 \\
23.75 &26816 & $\pm$   166 & 22764 & $\pm$   153 & 28190 & $\pm$   170 & 27884 & $\pm$   170 & 21319 & $\pm$   148 \\
24.25 &34314 & $\pm$   188 & 24761 & $\pm$   160 & 32020 & $\pm$   182 & 24579 & $\pm$   159 & 13479 & $\pm$   118 \\
24.75 &38195 & $\pm$   198 & 18709 & $\pm$   139 & 27257 & $\pm$   168 & 11529 & $\pm$   109 &  3946 & $\pm$    64 \\
25.25 &35118 & $\pm$   190 &  6659 &$\pm$    83 & 11820 & $\pm$   110 &  1774 & $\pm$    43 &   294 & $\pm$    17 \\
25.75 &23759 & $\pm$    156 &   929& $\pm$    31 &  1644 & $\pm$    41 &   121 & $\pm$    11 &    31 & $\pm$     6\\
\hline
\end{tabular}
\end{table*}

Fig.~\ref{fig:ncounts} shows the galaxy number counts for the VIDEO-XMM3 region compared to number counts derived from a variety of other near-infrared surveys. One can immediately see that in the $K_{\rm s}$-band the number counts agree with the number counts derived in this filter using the UltraVISTA survey \citep{McCracken2012} and the UKIDSS-DXS number counts from \citet{Kim2011}. Comparing the number counts with the WIRDS \citep[][]{Bielby2011} shows that the VIDEO number counts are also consistent\footnote{We use updated number counts supplied by R. Bileby (private communication)}. This is also true for the $J-$ and $H-$band data which are available as part of the WIRDS survey. 

We therefore conclude that there is strong agreement in the form of the galaxy number counts in VIDEO and other surveys. Table~\ref{tab:ncounts} provides the number counts (not corrected for completeness) within the VIDEO-XMM3 field for all filters.

 \section{Photometric Redshifts in the combined VIDEO and CFHT surveys}\label{sec:photozinvideo}
In this section we derive photometric redshift estimates for the combined VIDEO--CFHTLS-D1 optical survey using the publicly available photometric redshift code Le Phare\footnote{http://www.cfht.hawaii.edu/~arnouts/LEPHARE/lephare.html} \citep{Ilbert2006}. 
The VIDEO photometric redshifts were estimated using the optimised galaxy templates produced by \citet{Arnouts2007}. These are based on the four observed spectra in \citet{Coleman1980} an elliptical (Ell), two spirals (Sbc,Scd) and irregular galaxy (Irr) as well as two observed starburst (SB) templates from \citet{Kinney1996}. These templates have been linearly extrapolated to near-infrared and ultraviolet wavelengths using the GISSEL synthetic models \citep{BC2003} and optimised using data from  the VIMOS Very Large Telescope (VLT) Deep Survey (VVDS;~\citealt{lefevre2004,lefevre2005}), which targeted sources to $i_{\rm AB}<24.0$\,mag.

The process of constructing these optimised templates is described in detail in \citet{Ilbert2006}; they proceed by determining rest-frame spectral-energy distributions (SEDs) for all objects with secure redshifts and separating these into four categories based on the best-fit \citet{Coleman1980} template. Optimised templates are constructed from the median flux of the rest-frame spectra in each of these four categories; at wavelengths not sampled by the spectroscopic survey the original extrapolated \citet{BC2003} flux was retained. The final set of optimized templates is linearly interpolated to produce 64 reference templates used in the photometric-redshift fitting. For Scd and later spectral types the fitting also allowed for dust attenuation with permitted reddening excesses $E(B-V$) of 0.0, 0.05, 0.1, 0.15, 0.2 and 0.3\,mag. Dust attenuation was modelled using the interstellar extinction law measured by \citet{Prevot1984} in the Small Magellanic Cloud.  The opacity of the intergalactic medium is also included according to the prescription of \citet{Madau1995}. 

Redshifts were estimated from flux densities measured in 2~arcsec apertures and the associated errors were determined by measuring the rms noise in randomly placed apertures of the same size and adding Poisson noise. To adjust for uncertainties in the relative calibration of the observed bands additional uncertainties of 0.01\,mag were added in quadrature to the measured errors.

Systematic offsets in the photometric calibrations were also removed using a sample of approximately 3000 objects with spectroscopic redshifts in the VVDS-DEEP survey as described in \cite{Ilbert2006}. We note that the photometric offsets applied by Le Phare are not indicative of the uncertainty in the VIDEO photometry but are an indication of possible biases in the template set used. These offsets are listed in Table~\ref{tab:videooffset}.

To account for stars and quasars detected in the VIDEO survey the Le Phare code independently fits a set of stellar and AGN templates to the observed photometry and reports the results as $\chi_{star}$ and $\chi_{QSO}$. In the case of the quasar templates this process also reports the most likely redshift for a source with an intrinsic SED corresponding to the best fitting AGN template $z_{QSO}$. The AGN templates are those used and described in detail in \citet{Polletta2007}, the library consists of empirically determined spectra for three~Type~1 AGNs, five~Type~2 AGN and two composite AGN and starburst galaxies.

\begin{table}
\begin{center}
\caption{Systematic offsets (Mag$_{\rm data}$ - Mag$_{\rm template}$) for the ten filters used to estimate photometric redshifts, determined by calibrating against a sample of approximately 3000 objects with spectroscopic redshifts in the VVDS.}\label{tab:videooffset}
\begin{tabular}{llll}
\hline
Filter & offset [mag] & Filter & offset [mag]\\\hline
$u^*$ & 0.188& $Z$ & -0.037\\
$g'$ & -0.003 & $Y$ & -0.018\\
$r'$ & -0.003 & $J$ & 0.049\\
$i'$ & -0.059 & $H$ & 0.121\\
$z'$ & -0.064 &  $K_{\rm s}$ & 0.125 \\
\hline
\end{tabular}
\end{center}
\end{table}

\subsection{Comparison with Spectroscopic Redshifts}

\begin{figure}
\includegraphics[width=0.48\textwidth]{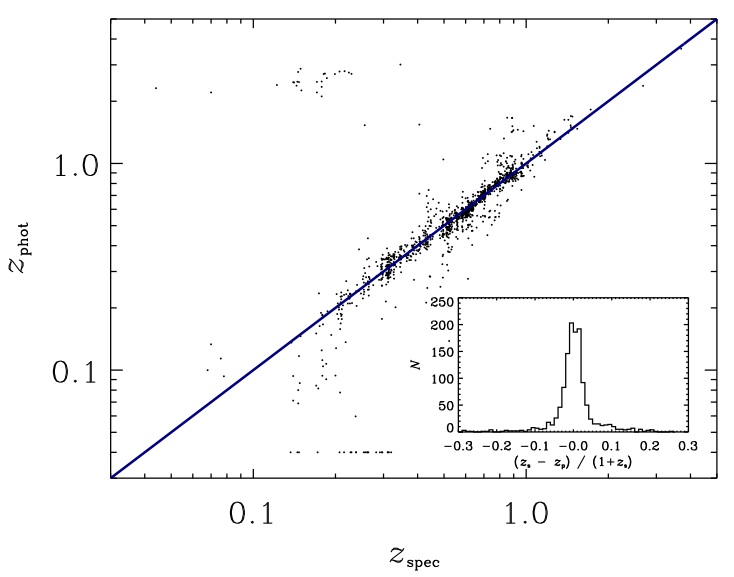}
\includegraphics[width=0.48\textwidth]{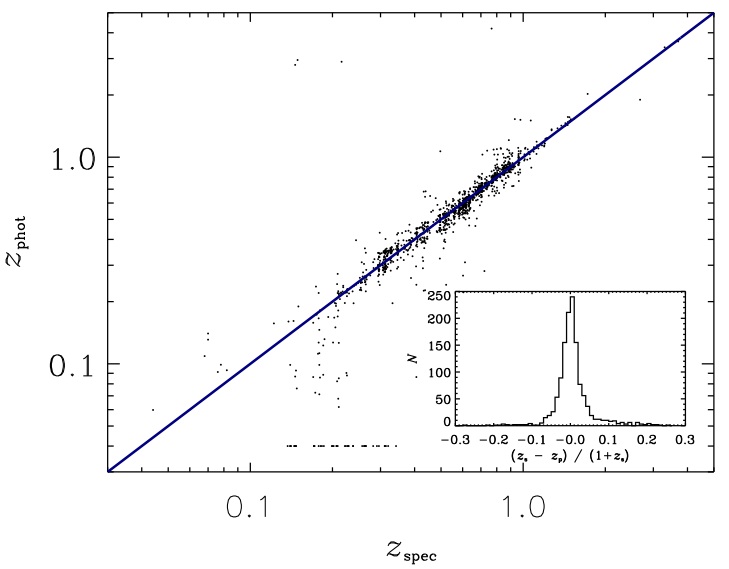}
\caption{Photometric redshifts versus high-reliability spectroscopic redshifts from the VVDS spectroscopic survey with photometric redshifts estimated without any luminosity function prior. {\it (Upper panel)} A comparison of the photometric redshift derived solely from the CFHTLS-D1 visible-wavelength filters ($ugriz$). {\it (Lower panel)} The same galaxies but where the photometric redshifts have been determined using the combined CFHTLS and $ZYJHK_{\rm s}$ VIDEO filters. The stripe of objects along the horizontal at $z=0.04$ is due to the photometric redshifts fitting to their lowest possible value. The inset shows the histogram of $\Delta z/(1+z)$.}
\label{fig:photoz}
\end{figure}

Photometric redshifts are known to produce `catastrophically' poor redshift estimates for some fraction of the sample. The `catastrophic' failure label refers to sources whose predicted redshifts differ from their spectroscopic redshifts by a margin considerably greater than the expected/predicted uncertainties. The most direct method of determining the accuracy of a photometric redshift procedure is by comparison with available spectroscopic data. In Fig.~\ref{fig:photoz} we present a comparison for galaxies with counterparts in the VVDS \citep{lefevre2005} using photometric redshifts determined from the CFHTLS visible-wavelength ($ugriz$) data alone and with the addition of the $ZYJHK_{\rm s}$ data from the VIDEO survey.


\begin{figure}
\includegraphics[width=0.48\textwidth]{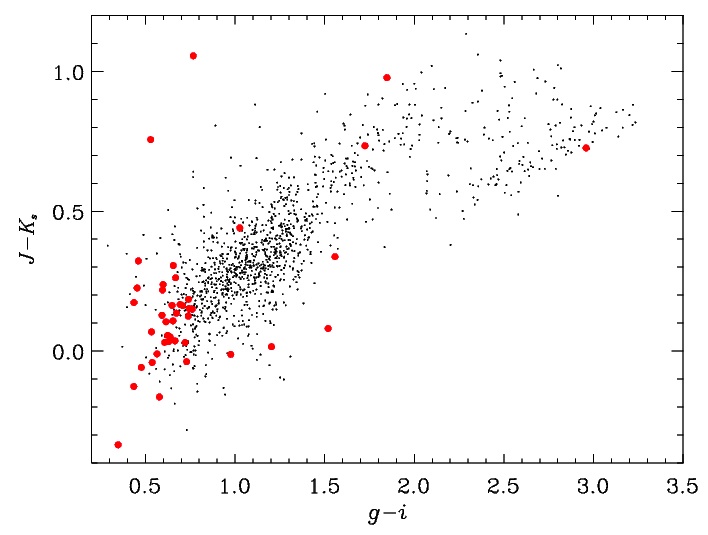}
\caption{$g-i$ versus $J-K_{s}$ colour-colour diagram showing the colours of the the VVDS spectroscopic sample used in our analysis The photometric-redshift outliers, defined as $|z_{s}-z_{p}|/(1+z_{s}) > 0.15$ are denoted by the filled red circles and the galaxies with good photometric redshifts, defined as $|z_{s}-z_{p}|/(1+z_{s}) \leq 0.15$, are shown as black dots. This shows that the bulk of the outliers have blue colours compared to the remainder of the galaxy population within the spectroscopic sample. }
\label{fig:outliers}
\end{figure}

Counterparts were identified by cross-matching the VIDEO and VVDS source positions to within 1~arcsec tolerance. The VVDS catalogue provides quality flags indicating confidence in the reported redshift. Quality flags 2,~3 and 4 indicate confidence levels of 75,~95 and 100 per cent respectively whilst objects with a flag of 1 have only indicative redshifts based on the continuum with very few supporting spectral features ($\sim$ 50\% confidence). The least secure redshifts are those with quality flag 9 which are based on only a single emission line and assigned a tentative redshift. To evaluate the accuracy of the VIDEO photometric redshifts we use only those sources in the VVDS sample with quality flags of 4. We also remove objects classified as stars by the colour selection criteria described in section~\ref{sec:numcounts} and sources classified as quasars in the VVDS spectroscopy. Following \cite{Ilbert2006}, we determine the normalized median absolute deviation \citep[NMAD; ][]{Hoaglin1983} in $\Delta z/(1+z_{s})$, where $\Delta z=z_{s}-z_{p}$ with $z_{s}$ and $z_{p}$ denoting the spectroscopic and photometric redshifts respectively. For the VVDS sample we find the NMAD in $\Delta z/(1+z_{s})$ to be $\sigma = 0.025$ and $\sigma=0.023$ for the CFHTLS and CFHTLS+VIDEO data sets respectively. However, the fraction of outliers, defined as sources with $|z_{s}-z_{p}|/(1+z_{s}) > 0.15$ reduced from 7.4 per cent for the CFHTLS filters alone to  3.8 per cent when the VIDEO filters are added, if no prior on the luminosity function is used \citep[c.f.][]{Ilbert2006}. If we use the $i-$band luminosity function prior, as detailed in \cite{Ilbert2006}, we find that the fraction of outliers with just the CFHT data is significantly reduced to 5.4 per cent, whereas implementing the prior with the VIDEO data included reduces the fraction of outliers to 3.3 per cent, although a reduced outlier rate of 2 per cent can be achieved by adopting a liberal $\chi^2$ cut. Therefore, even at $z<1$ where the majority of the spectroscopically determined redshifts lie, the VIDEO data reduce the number of catastrophic outliers significantly.

We are able to investigate whether the outliers occupy certain areas of colour-colour space in which the photometric redshifts do not perform very well. Fig.~\ref{fig:outliers} shows the $g-i$ versus $J-K_{\rm s}$ colour-colour diagram for the VVDS sample with good redshift quality. This shows that the galaxies with poor photometric redshifts are generally blue compared to the larger sample with good photometric redshifts. This is unsurprising as blue galaxies do not exhibit strong spectral breaks at  low redshift and are therefore difficult to accurately fit photometric redshifts. Inspecting the individual spectra, we also find a significant fraction with very bright emission lines and although emission lines are incorporated into the photometric redshift templates, they are unable to accurately reproduce all emission line ratios found in galaxies. Indeed, the emission line ratios in some of the galaxies appear to indicate the presence of an AGN. To investigate this further,  we restrict the sample to only those galaxies with a best fitting galaxy template, rather than include galaxies which are better fit by an AGN template.  We find that the percentage of photometric-redshift outliers in this case to decrease further, to 1.2 and 0.3 per cent for the CFHT and CFHT+VIDEO photometry, respectively.

As noted above, there is a dearth of spectroscopic redshifts in the $1<z<4$ redshift range which is the niche redshift range for the VIDEO survey. We can expect that the photometric accuracy for galaxies at these redshifts will increase substantially with the near-infrared data provided by the VIDEO survey, as the Balmer and 4000\AA\, breaks are redshifted into the near-infrared filters \citep[e.g.][]{Abdalla2008, Ilbert2009}.

\begin{figure}
\includegraphics[width=0.48\textwidth]{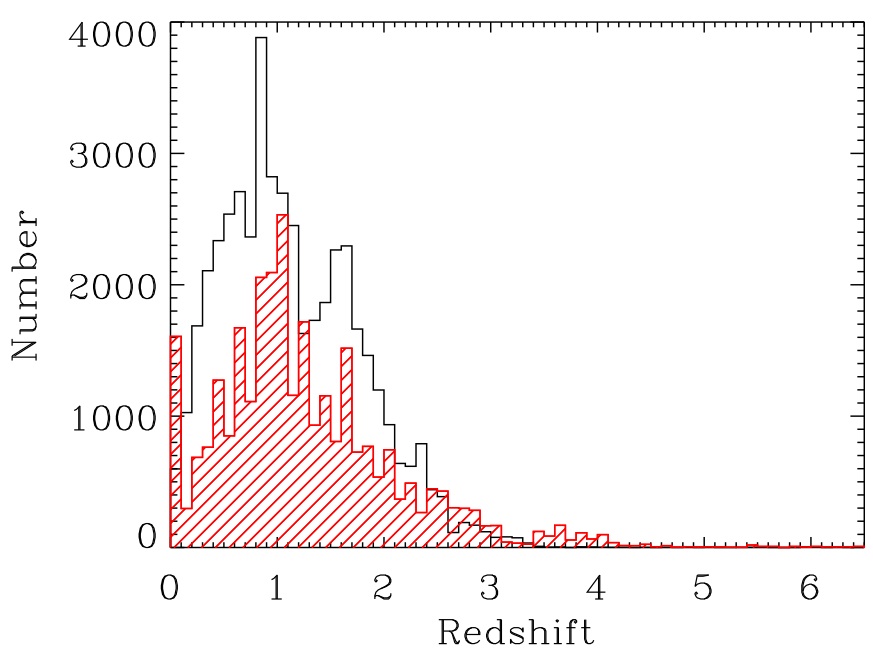}
\caption{Photometric redshift distribution for all sources in the VIDEO-XMM3-CFHTLS-D1 overlap area with $K_{\rm s} < 23.5$ The solid black line is for all sources classified as galaxies from the photometric redshift fitting with Le Phare. The red hashed region shows the distribution of sources classified as AGN by Le Phare (see text for details).}
\label{fig:dndz}
\end{figure}

In Fig.~\ref{fig:dndz} we show the redshift distribution for all sources detected in the VIDEO to a limit of $K_{\rm s} \le\,23.5$. It is immediately apparent that VIDEO provides an excellent data set to study galaxy evolution over cosmic time with the redshift distribution peaking at $z\sim\,1$ and extending all the way through to $z\sim\,6$. The peak observed at $0.8<z<0.9$ indicates the presence of a large structure of galaxies at this redshift, although further work is needed to confirm whether it is indeed real.
Also shown in Fig.~\ref{fig:dndz} is the photometric redshift distribution for sources classified as AGN by Le Phare, taken to be those objects with a lower $\chi^2$ for a QSO template than for a galaxy template. The redshift distribution of AGN tends to extend to higher redshifts than those sources classified by galaxies, as we would expect, given that on average they are brighter than the quiescent galaxy population.

\section{Summary}

In this paper we have given an overview of the scientific goals, the data processing and the technical details of the VISTA Deep Extragalactic Observations (VIDEO) survey.  Our stacked images from our first major data release in the VIDEO-XMM3 field reach $5\sigma$ depths of $Z=25.7$, $Y=24.5$, $J=24.4$, $H=24.1$ and $K_{\rm s}=23.8$ in the AB-mag system in a 2~arcsec diameter aperture, with the expectation that the full depth in the $Y$ band will be reached in the planned integration times. This quality of data will eventually cover 12~degree$^2$ of the best-studied fields in the southern hemisphere, making it a unique survey for investigating the formation and evolution of the rarest and most massive galaxies down to the sub-$L^{\star}$ galaxy population as a function of galaxy environment, over 90 per cent of cosmic time.

For this first major data release, we have emphasized the power of combining the VIDEO survey with the CFHTLS-D1 optical data in the VIDEO-XMM3 field for improving photometric redshifts (these can be obtained from the VIDEO consortium web page\footnote{http://star.herts.ac.uk/$\sim$dgb/video}). The addition of the VIDEO bands to optical data results in a considerable decrease in the fraction of catastrophic failures in the photometric redshifts, from 7.4 per cent to 3.8 per cent without an $i-$band luminosity function prior and from 5.4 to 3.3 per cent when the luminosity function prior is included. Furthermore, the accuracy of the photometric redshifts with the inclusion of the near-infrared VIDEO data should increase substantially at $z>1$ due to the ability to bracket strong spectral features, such as the Balmer and 4000\AA\, breaks, over a broader range of redshifts. However, we also emphasize the importance of additional deep optical imaging data over the full VIDEO Survey area. Thus, future observations as part of the Dark Energy Survey and the VLT Survey Telescope (VST) will play a crucial role in the full scientific exploitation of the VIDEO data.

The images and catalogues described here will be publicly available from the ESO archive\footnote{http://www.eso.org/sci/observing/phase3/data\_releases/} and the VISTA Science Archive\footnote{http://surveys.roe.ac.uk/vsa}. Access to value-added information, and updates on the progress of the VIDEO survey can also be found at the VIDEO survey website (http://star.herts.ac.uk/$\sim$dgb/video). The forthcoming release at the end of June 2012 will also include $Z$ and $K_{\rm s}-$band data in the ELAIS-S1 northern field, full details of which will accompany the data within the ESO archive and the VSA.

The remainder of the VIDEO Survey data and value-added catalogues will be released on regular 6-12 month timescales. A description of the data products will accompany all releases and be held at the ESO and VSA archives, as well as on the VIDEO survey website.

\section*{ACKNOWLEDGEMENTS} 
This paper is dedicated to the memory of Steve Rawlings, without whom the VIDEO Survey may not have gotten off the ground, but also for the never ending enthusiasm, encouragement and occasional drink during 15 years working with MJJ.

MJJ and DGB acknowledges the support from STFC through the University of Hertfordshire Extragalactic Rolling Grant and Consolidated Grant. VAB and SVW acknowedge support of STFC studentships and RJM would like to acknowledge the funding of the Royal Society via the award of a University Research Fellowship and the Leverhulme Trust via the award of a Philip Leverhulme research prize.
JSD acknowledges the support of the European Research Council through an Adanced Grant, and the support of the Royal Society via a Wolfson Research Merit Award. MK acknowledges the support from the Creative Research Initiative program, No. 2010-0000712, of the National Research Foundation of Korea (NRFK) funded by the Korea government(MEST). MK was partially supported by the EU COST Action “Black Holes in a Violent Universe”. SJO acknowledges support from the Science and Technology Facilities Council [grant number ST/I000976/1]. IS acknowledges support from STFC and a Leverhulme Senior Fellowship.

The CFHT data used in this paper were based on observations obtained with MegaPrime/MegaCam, a joint project of CFHT and CEA/DAPNIA, at the Canada-France-Hawaii Telescope (CFHT) which is operated by the National Research Council (NRC) of Canada, the Institut National des Science de l'Univers of the Centre National de la Recherche Scientifique (CNRS) of France, and the University of Hawaii. This work is based in part on data products produced at TERAPIX and the Canadian Astronomy Data Centre as part of the Canada-France-Hawaii Telescope Legacy Survey, a collaborative project of NRC and CNRS.
{} 

\end{document}